\documentclass[pdftex,cite,aps,prl,superscriptaddress,amsmath,amsfonts,amssymb,twocolumn,showpacs]{revtex4}
\usepackage{wasysym}
\renewcommand{\vec}[1]{\mathbf{#1}} 
\usepackage[pdftex]{graphicx}
\usepackage{times}

\newcommand{\figref}[1]{Fig.~\ref{fig:#1}}
\newcommand{\Figref}[1]{Figure~\ref{fig:#1}}

\renewcommand{\eqref}[1]{Eq.~\ref{eq:#1}}

\newcommand{\citeasnoun}[1]{Ref.~\onlinecite{#1}}

\begin{document}

\title{Geometry-induced Casimir suspension of oblate bodies in fluids}

\author{Alejandro W. Rodriguez}
\affiliation{Department of Electrical Engineering, Princeton University, Princeton, NJ 08540, USA}
\author{M. T. Homer Reid}
\affiliation{Department of Mathematics, Massachusetts Institute of Technology, Cambridge, MA 02139, USA}
\author{Francesco Intravaia}
\affiliation{Theoretical Division, MS B213, Los Alamos National Laboratory, Los Alamos, NM 87545, USA}
\affiliation{School of Physics and Astronomy, The University of Nottingham, University Park, NG7 2RD Nottingham, United Kingdom}
\author{Alexander Woolf}
\affiliation{School of Engineering and Applied Sciences, Harvard University, Cambridge, MA 02138, USA}
\author{Diego~A.~R. Dalvit}
\affiliation{Theoretical Division, MS B213, Los Alamos National Laboratory, Los Alamos, NM 87545, USA}
\author{Federico Capasso}
\affiliation{School of Engineering and Applied Sciences, Harvard University, Cambridge, MA 02138, USA}
\author{Steven G. Johnson}
\affiliation{Department of Mathematics, Massachusetts Institute of Technology, Cambridge, MA 02139, USA}

\begin{abstract}
  We predict that a low-permittivity oblate body (disk-shaped object)
  above a thin metal substrate (plate with a hole) immersed in a fluid
  of intermediate permittivity will experience a meta-stable
  equilibrium (restoring force) near the center of the hole. Stability
  is the result of a geometry-induced transition in the sign of the
  force, from repulsive to attractive, that occurs as the disk
  approaches the hole---in planar or nearly-planar geometries, the
  same material combination yields a repulsive force at all
  separations in accordance with the
  Dzyaloshinski{\u{\i}}--Lifshitz--Pitaevski{\u{\i}} condition of
  fluid-induced repulsion between planar
  bodies~\cite{Dzyaloshinskii61}.  We explore the stability of the
  system with respect to rotations and lateral translations of the
  disks, and demonstrate interesting transitions (bifurcations) in the
  rotational stability of the disks as a function of their
  size. Finally, we consider the reciprocal situation in which the
  disk--plate materials are interchanged, and find that in this case
  the system also exhibits meta-stability. The forces in the system
  are sufficiently large to be observed in experiments and should
  enable measurements based on the diffusion dynamics of the suspended
  bodies.
 \end{abstract}

\maketitle 

Casimir forces arising from quantum/thermal fluctuations of charges
are becoming increasingly important in nano- and micro-scale
systems~\cite{casimir,milton04,Lamoreaux07:phystoday,Genet08,Bordag09:book,Klimchitskaya09,Rodriguez11:review,Dalvit11:review},
where the usually attractive nature of the force leads to unwanted
effects such as stiction~\cite{Rodriguez11:review}.  Recent
theoretical developments have made it possible to study the influence
of geometry and materials on these
interactions~\cite{Johnson11:review,Rodriguez11:review}; for instance,
geometry effects alone can lead to unusual behaviors, including
non-monotonic and/or repulsive forces between vacuum-separated
bodies~\cite{Rodriguez07:PRL,Rodriguez-Lopez09,LevinMc10}. For
planar geometries, one way to obtain repulsion is to employ
fluids~\cite{isra,Dzyaloshinskii61,Parsegian06}. Dzyaloshinski{\u{\i}}
\emph{et. al.} showed decades ago that two planar bodies of
permittivities $\varepsilon_{1,2}$ immersed in a fluid of permittivity
$\varepsilon_3$, satisfying $\varepsilon_1 < \varepsilon_3 <
\varepsilon_2$ will repel one another~\cite{Dzyaloshinskii61}, an
effect that has also been observed in
experiments~\cite{Feiler08,Munday09}. Based on that prediction, one
might ask whether the
Dzyaloshinski{\u{\i}}--Lifshitz--Pitaevski{\u{\i}} (DLP) condition
alone suffices to obtain repulsion regardless of geometry. In this
letter, we exploit a recently developed numerical method for computing
Casimir interactions between arbitrary bodies~\cite{Reid11} to answer
this question in the negative. Specifically, we show that the Casimir
potential between an oblate body (a disk-shaped object) and a thin
metal substrate (a plate with a hole) immersed in a fluid satisfying
the DLP condition exhibits a meta-stable equilibrium at the center of
the hole, creating a ``Casimir trap'' for the disk.

\begin{figure}[t!]
\includegraphics[width=1.0\columnwidth]{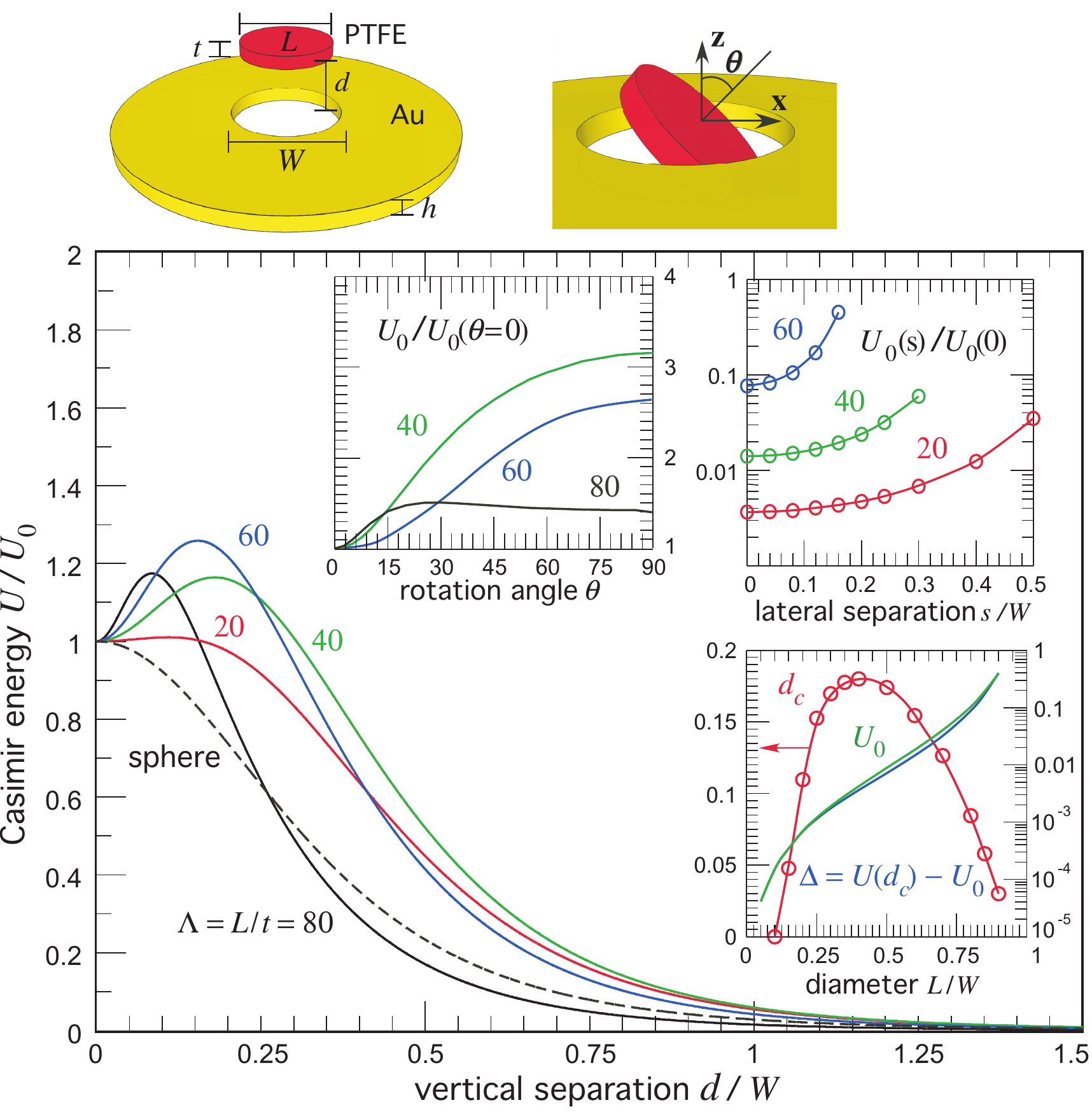}
\caption{Room-temperature Casimir energy $U$ of a PTFE disk (thickness
  $t=10$nm) suspended above a Au plate (thickness $h=10$nm and inner
  and outer diameters $W=1\mu$m and $D=2W$) immersed in water, as a
  function of vertical separation $d$ (normalized by $W$). $U$ is
  normalized by the energy in the co-planar configuration, $U_0\equiv
  U(d=0)$, and plotted for multiple aspect ratios $\Lambda=L/t$, where
  $L$ is the disk diameter. Also shown is the energy of a sphere of
  diameter $20t$ (dashed black line).  (Top insets:) $U$ as a function
  of rotation angle $\theta$ (left) and lateral translations $s$
  (right) for multiple $\Lambda$. (Bottom inset:) unstable equilibrium
  separation $d_c$ (red circles), along with the energy $U_0$ (green
  line) and corresponding energy barrier $\Delta = U(d_c)-U_0$ (blue
  line), normalized by $k_B T\approx 25$m$eV$, as a function of
  $L$. Both $d_c$ and $L$ are normalized by $W$.}
\label{fig:fig1}
\end{figure}

Although Casimir suspensions are impossible for vacuum-separated
bodies (irrespective of geometry)~\cite{KennethKl06,Rahi10:PRL}, they
can arise in fluids satisfying the DLP
condition~\cite{Rodriguez08:PRL,McCauleyRo10:PRA,RodriguezMc10:PRL}. The
approach described here differs from previous work in that it does not
rely on material dispersion~\cite{RodriguezMc10:PRL} or the presence
of external forces (e.g. gravity~\cite{McCauleyRo10:PRA}), nor does it
require bodies to be enclosed inside one
another~\cite{Rodriguez08:PRL,RahiZa10}, but instead stems from the
anomalous behavior of electromagnetic fields in this particular
geometry (shown schematically in \figref{fig1}).  In recent
work~\cite{LevinMc10}, we exploited a similar geometric effect to
demonstrate the possibility of switching the sign of the Casimir force
between two \emph{vacuum}-separated bodies---a small, metallic,
\emph{prolate body} (thin needle) centered above a metal plate with a
hole---from attractive to \emph{repulsive}. That phenomenon was
explained via a simple symmetry argument~\cite{LevinMc10}: because the
fields of a needle in vacuum behave like those of a dipole oriented
along its symmetry axis, its interaction with a plate decreases as the
needle reaches the center of the hole (at which point the field lines
become orthogonal to the plate). The same symmetry argument (in
conjunction with a more sophisticated dipole model) is employed here
to show that in water, the interaction between a small
Polutetraflouroethylene (PTFE) body and a thin gold (Au) plate with a
hole can be switched from repulsive to \emph{attractive} near the
vicinity of the hole. To our surprise, however, the interesting
geometry in this case is not a needle but rather an \emph{oblate body}
(thin disk), a consequence of the flipped polarization-response of the
disk in the fluid. We quantify deviations from dipole-like behavior by
comparing our Casimir predictions against a corresponding
Casimir--Polder (CP) model in which the disk is modelled as a dipole
of equivalent polarizability, and show that finite-size effects can
lead to significant qualitative and quantitative deviations for large
disks and small separations. Interestingly, despite these deviations,
we find that the desired geometric effects persist even for
\emph{large} disks (with diameters $\sim$ hole size), leading to much
larger forces than those predicted in the vacuum case. Moreover,
unlike the vacuum case (in which the needle must be anchored to a
static surface~\cite{LevinMc10,McCauleyRo11}), here the disks are
stable with respect to rotations and/or lateral translations, and are
therefore free to move subject to Brownian motion. This enables
exploration of this phenomenon through a broader set of experimental
techniques, e.g. measurements based on total-internal reflection
microscopy or diffusion dynamics. Finally, we consider the
``reciprocal'' situation involving a Au disk above a PTFE plate, and
find that in that case one also obtains a meta-stable equilibrium,
albeit with larger geometric anisotropy leading to larger energy
barriers.

\Figref{fig1} shows the room-temperature Casimir energy $U$ between a
PTFE disk and a co-axial Au plate immersed in water, as a function of
their mutual center--center separation $d$. $U$ is normalized by the
energy $U_0$ when the two bodies are coplanar ($d=0$) and is plotted
for multiple aspect ratios $\Lambda=L/t$ (keeping $t$ fixed). The Au
dielectric permittivity is obtained from a Drude model with plasma
frequency $\omega_\mathrm{p}=9$eV and damping constant
$\gamma=0.035$eV, whereas the PTFE and water permittivities are
obtained using the oscillator models described
in~\citeasnoun{Zwol10}. This specific material combination was chosen
because it satisfies the DLP condition of fluid repulsion between
planar bodies---indeed, we find that the force between a finite disk
and an unpatterned ($W=0$) plate is repulsive over all $d$ and
diverges as $d\to 0$ (not shown). As expected, and in contrast to the
unpatterned case, the presence of the hole means that $U$ no longer
diverges as $d\to 0$ but instead reaches a finite constant (so long as
$L < W$). We find that for spheres (dashed black line),
nearly-isotropic, or prolate bodies, $U$ increases monotonically with
decreasing $d$, attaining its peak at $d=0$ as expected. The situation
is different for oblate bodies ($\Lambda > 1$), in which case $U$
peaks at a critical separation $d_c > 0$ (determined by $\Lambda$),
below which the force transitions from repulsive to
\emph{attractive}. In particular, instead of the usual unstable
equilibrium, we find that the disk exhibits a \emph{meta-stable}
equilibrium at $d=0$. In order to investigate the full stability of
the disk, and its dependence on $\Lambda$, the top insets in
\figref{fig1} show the energy of the system in the co-planar
configuration ($d=0$) as a function of rotation $\theta$ and lateral
translations $s$ of the disk, for multiple $\Lambda$.  Our results
reveal that whenever $\Lambda$ is either too small or too large, the
non-monotonicity in the potential (and corresponding meta-stability)
disappears. Specifically, we find that $d_c$ and the corresponding
potential barrier $\Delta=U(d_c)-U_0$ vanish as $L \to 0$ and $L\to W$
(not shown in the figure), respectively. Moreover, while the disk is
repelled from the edges of the hole irrespective of $\Lambda$, its
stability with respect to rotations changes drastically with
increasing $L/W$. In particular, beyond $L \approx 0.7 W$,
corresponding to $\Lambda\approx 80$, additional unstable and stable
equilibria appear at (a finite) $\theta_c > 0$ and $\theta=90^\circ$,
respectively.  For $L \gtrsim 0.9W$ (not shown), corresponding to
$\Lambda \approx 90$, the preferred orientation of the disk (the
minimum $U$) changes from $\theta=0$ (parallel) to $\theta=90^\circ$
(perpendicular). In the perpendicular orientation, the potential
barrier $\Delta \to 0$ and the disk is repelled from the hole.

\begin{figure}[t!]
\includegraphics[width=1.0\columnwidth]{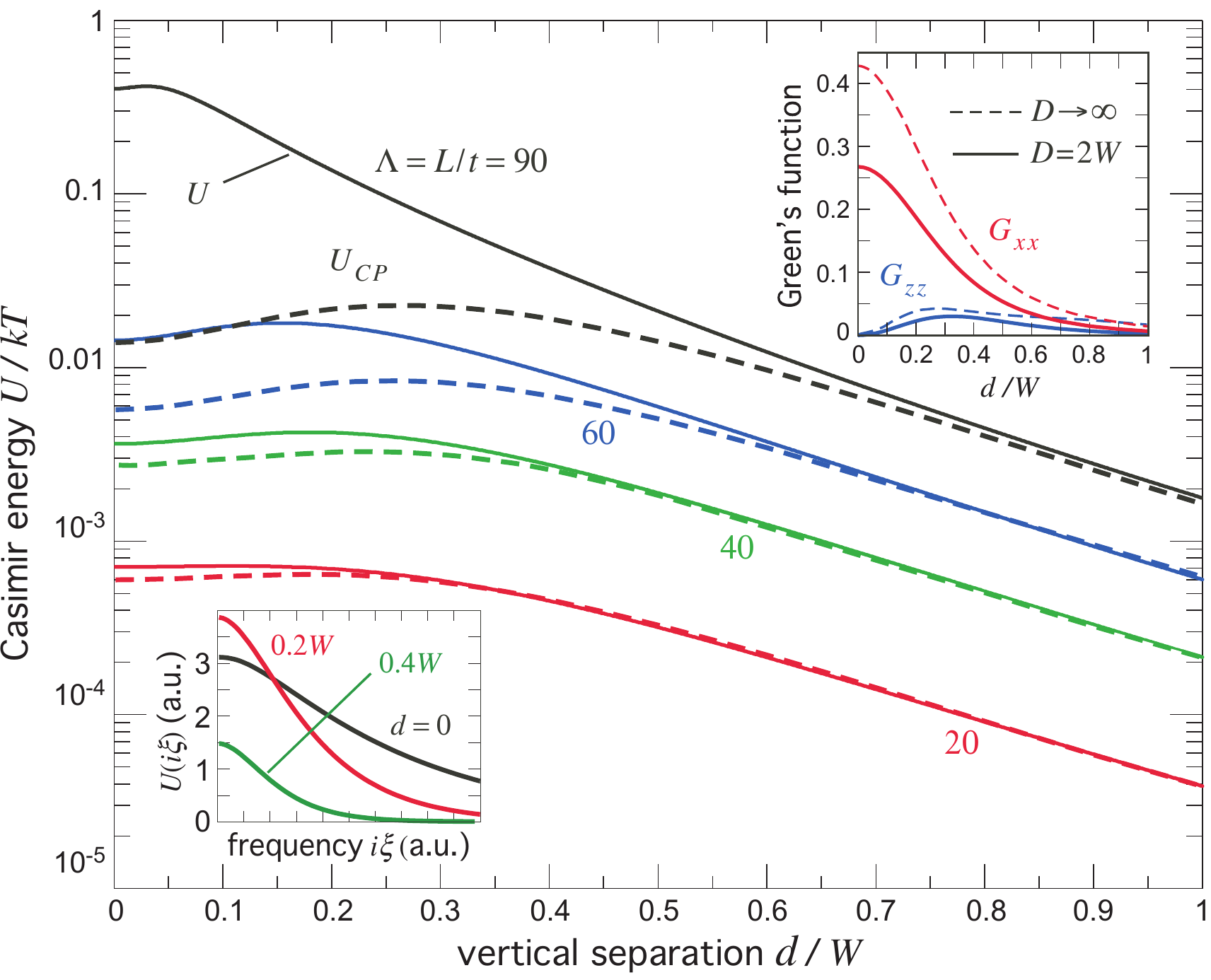}
\caption{Room-temperature Casimir $U$ (solid lines) and
  Casimir--Polder $U_{CP}$ (dashed lines) energies, normalized by $k_B
  T\approx 25$m$eV$, for the disk--plate geometry of \figref{fig1}, as
  a function of $d$ (normalized by $W$), plotted for multiple
  aspect-ratios $\Lambda=L/t$.  $U_{CP}$ is obtained from the
  polarizability of the disks, as determined by \eqref{alpha}. (Top
  inset:) diagonal components of the photon GF of the plate evaluated
  along the axis of symmetry, for both finite $D=2W$ (solid lines,
  evaluated numerically) and semi-infinite $D \to \infty$ PEC (dashed
  lines, evaluated anaytically~\cite{Eberlein11}) plates. (Bottom
  inset:) Casimir integrand $U(i\xi)$, in arbitrary units, as a
  function of imaginary frequency $i\xi$ at three different
  separations $d=\{0, 0.2, 0.4\}W$, for $\Lambda=60$.}
\label{fig:fig2}
\end{figure}

In order to understand the above features as well as the origin of the
non-monotonicity in $U$, it is useful to examine the Casimir-energy
imaginary-frequency spectrum $U(i\xi$) of the system
, whose integral (a Matsubara sum at finite
temperatures~\cite{Rodriguez11:review}) yields $U$. The bottom inset
of \figref{fig2} shows $U(i\xi)$ for a representative disk--plate
configuration exhibiting non-monotonicity ($\Lambda=60$) at multiple
separations $d=\{0,0.2,0.4\}W$, and illustrates that non-monotonicity
in $d$ is present only at small ``quasistatic'' $\xi$. 
In this quasistatic regime, a thin disk immersed in a
fluid of larger permittivity will act like a fluctuating dipole
oriented mainly along its symmetry axis~\footnote{Strictly speaking,
  the dipole limit requires $L \ll W \ll d$.}. In contrast, the same
disk in vacuum will be mainly polarized in the direction
\emph{transverse} its axis of symmetry (as shown below). Since the
fields generated by a fluctuating dipole lie mainly along the dipole
axis and become orthogonal to the metal plate as $d\to 0$, it follows
that the disk--plate interaction will weaken in the vicinity of the
hole~\cite{LevinMc10}, leading to the behavior above.

In what follows, we quantify the previous argument via a simple model
in which the disk is described as a dipole with an effective
polarizability, corresponding to the leading-order term of a
spherical-harmonic expansion in the scattering
formalism~\cite{Emig09:ellipsoids}. The zero-temperature CP energy
between a polarizable particle at position $\vec{x}$ and the plate can
be written as~\cite{Casimir48:polder,Dalvit11:review}:
\begin{equation}
  U_{CP} = -\frac{\hbar}{2\pi} \, \int_0^\infty d\xi\, \, \mathrm{Tr} \left[\boldsymbol{\alpha}(i\xi) \cdot \mathbb{G}(i\xi;
    \vec{x},\vec{x})\right],
\label{eq:UCP}
\end{equation}
where $\boldsymbol{\alpha}(i\xi)$ and
$\mathbb{G}(i\xi,\vec{x},\vec{x})$ are the imaginary-frequency dipole
polarizability and Dyadic Green's function (GF) of the plate in the
surrounding medium evaluated at the location of the dipole.

Although the polarizability of a disk of permittivity $\varepsilon_1$,
diameter $L$, height $t$, and corresponding volume $V=\pi L^2 t$,
surrounded by a medium of permittivity $\varepsilon_3$, cannot be
easily computed analytically, it is nevertheless well approximated by
that of a spheroidal body of similar
dimensions~\cite{Hulst81,Venermo05}. In that case, the polarizability
in the $j$th direction,
\begin{equation}
  \alpha_j = \varepsilon_3^2 V \frac{\tau - 1}{1 + (\tau-1) n_j},
\label{eq:alpha}
\end{equation}
is determined by the ratio $\tau \equiv \varepsilon_1 / \varepsilon_3$
and depolarization factors, $n_{x,y} = \frac{1}{2} \left(1-n_z\right)$
and
\begin{align}
  n_z &= \begin{cases}
    \frac{1-e^2}{2e^3} \left[\log\left(\frac{1+e}{1-e}\right)-2e \right], & t < L \\
    \frac{1+e^2}{e^3} \left[e - \arctan(e)\right], & t > L,
    \end{cases}
\end{align}
where $e = \sqrt{|1 - (t/L)^2|}$ is the eccentricity of the
body~\cite{Venermo05}.

To qualitatively explain the behavior observed in \figref{fig1}, it
suffices to restrict our analysis to the asymptotic limits of either
an elongated ``needle'' (a prolate body with $t \gg L$) or a flat
``disk'' (an oblate body with $t \ll L$), in which case
\begin{align}
  \alpha_z = \varepsilon_3^2 V
  \begin{cases}
    \tau-1, & t \gg L \\
    1-\frac{1}{\tau}, & t \ll L
  \end{cases},
  \hspace{0.1in} 
  \alpha_{x,y} = \varepsilon_3^2 V
  \begin{cases}
    2\frac{\tau-1}{\tau+1}, & t \gg L \\
    \tau-1, & t \ll L
  \end{cases}.
\end{align}
Matters simplify further in the limit of large index
contrast ($\tau \ll 1$ or $\tau \gg 1$), in which case the CP energies
of the needle $U_{needle}$ and disk $U_{disk}$ take the form:
\begin{align}
  \label{eq:Uneedle}
  U_{needle} &= -\varepsilon_3^2 V
  \begin{cases}
    \tau G_{zz}, & \tau \gg 1 \\
    -G_{zz} - 2 (G_{xx}+G_{yy}), & \tau \ll 1
  \end{cases} 
  \\
  U_{disk} &=
  -\varepsilon_3^2 V
  \begin{cases}
    \tau (G_{xx} + G_{yy}), & \tau \gg 1 \\
    -\frac{1}{\tau} G_{zz}, & \tau \ll 1,
  \end{cases}
  \label{eq:Udisk}
\end{align}
where $G_{kk} \equiv G_{kk}(i\xi; \vec{x},\vec{x})$. In the case of an
infinitesimally thin perfect electric conductor (PEC) plate
(corresponding to $\varepsilon_2 \to \infty$, $h\to 0$, and $D\to
\infty$ in our geometry) with a hole of size $W$, one can write down
an analytical expression for $G_{kk}(0; \vec{x},\vec{x})$ in the
non-retarded limit~\cite{Eberlein11,Milton11}. For a dipole centered
along the axis of symmetry of the plate, i.e. $\vec{x}=\{0,0,d\}$, one
finds that $G_{zz}$ exhibits local minima and maxima at $d=0$ and
$d=d_c\approx W/3$, respectively, while $G_{xx}=G_{yy} (\gg G_{zz})$
decreases monotonically with separation~\cite{Eberlein11}. Both GF
components are plotted versus $d$ on the top inset of \figref{fig2}
(dashed lines). It follows from \eqref{Uneedle} that a needle will
experience a repulsive (attractive) force for $d < d_c$ ($d > d_c$) in
the $\tau \gg 1$ regime, as was predicted in~\citeasnoun{LevinMc10},
and a repulsive force at all separations in the $\tau \ll 1$ regime,
in agreement with predictions based on the DLP condition. In contrast,
however, \eqref{Udisk} predicts that a disk will experience an
attractive force at all separations in the $\tau \gg 1$ regime, and an
attractive (repulsive) force for $d < d_c$ ($d > d_c$) in the $\tau
\ll 1$ regime, in qualitative agreement with our results above.

In order to incorporate effects coming from the finite size/thickness
of the plate, as well as to quantify deviations from the dipole
picture that arise in the $L \to W$ and $d\to 0$ limits, we compare
our results of \figref{fig1} to the corresponding CP potential of the
system, obtained via \eqref{UCP} by assuming a spheroidal particle
with polarizability given by \eqref{alpha} and with $\mathbb{G}(i\xi)$
computed numerically. As expected, the GFs of the finite Au plate,
plotted in the $\xi \to 0$ limit on the top inset of \figref{fig2}
(solid lines), are smaller than those of the semi-infinite PEC plate
(due to its smaller surface area), but exhibit the same anomalous
behavior. (Away from the quasistatic regime, corresponding to larger
$\xi$, $G_{zz}$ exhibits non-monotonicity but $\tau$ tends to unity,
causing the object to appear more isotropic as can be seen from
\eqref{alpha}, and leading to the dissapearance of this effect.)
\Figref{fig2} shows both $U_{CP}$ and the Casimir energy $U$ versus
$d$, for multiple values of $\Lambda$ (with $t$ fixed as before),
showing agreement at large $d$ and small $\Lambda$, a regime where the
disks behave like ideal (isolated) dipoles. In the opposite limit of
large $\Lambda \gg 1$ (corresponding to $L \sim W$), the outer and
inner surfaces of the disk and plate approach one another (touching as
$d\to 0$ for $L \geq W$), thereby causing the interaction energy to be
dominated by proximity effects~\cite{Dalvit11:review}. This transition
manifests itself in multiple ways: First, though the CP model predicts
a monotonically increasing $d_c$ with increasing $L$, we find instead
that in the finite system, $d_c$ reaches a maximum at $L\approx 0.4W$
and then \emph{decreases} as $L\to W$ (red circles on the bottom inset
of \figref{fig1}). Second, while the dipole picture predicts a
monotonically increasing $U_{CP} \sim L^2$ (stemming from the linear
dependence of the polarizability with the disk volume), the dependence
of $U_0$ on $L$ exhibits a power-law \emph{divergence} that scales as
$1/(W-L)^{\beta}$, with $\beta \approx 5/2$, in the limit as $L \to W$
(green line on the bottom inset of \figref{fig1}).  The same proximity
effects are responsible for a dramatic increase in $\Delta$ (blue
line) with increasing $L$. We note however, that the competition
between increasing $U$ and decreasing nonmonotonicity eventually skews
in favor of the latter causing a peak in $\Delta$ as $L \to W$ and
eventually causing $\Delta \to 0$ in this limit (not shown in the
figure).

\begin{figure}[t!]
\includegraphics[width=1.0\columnwidth]{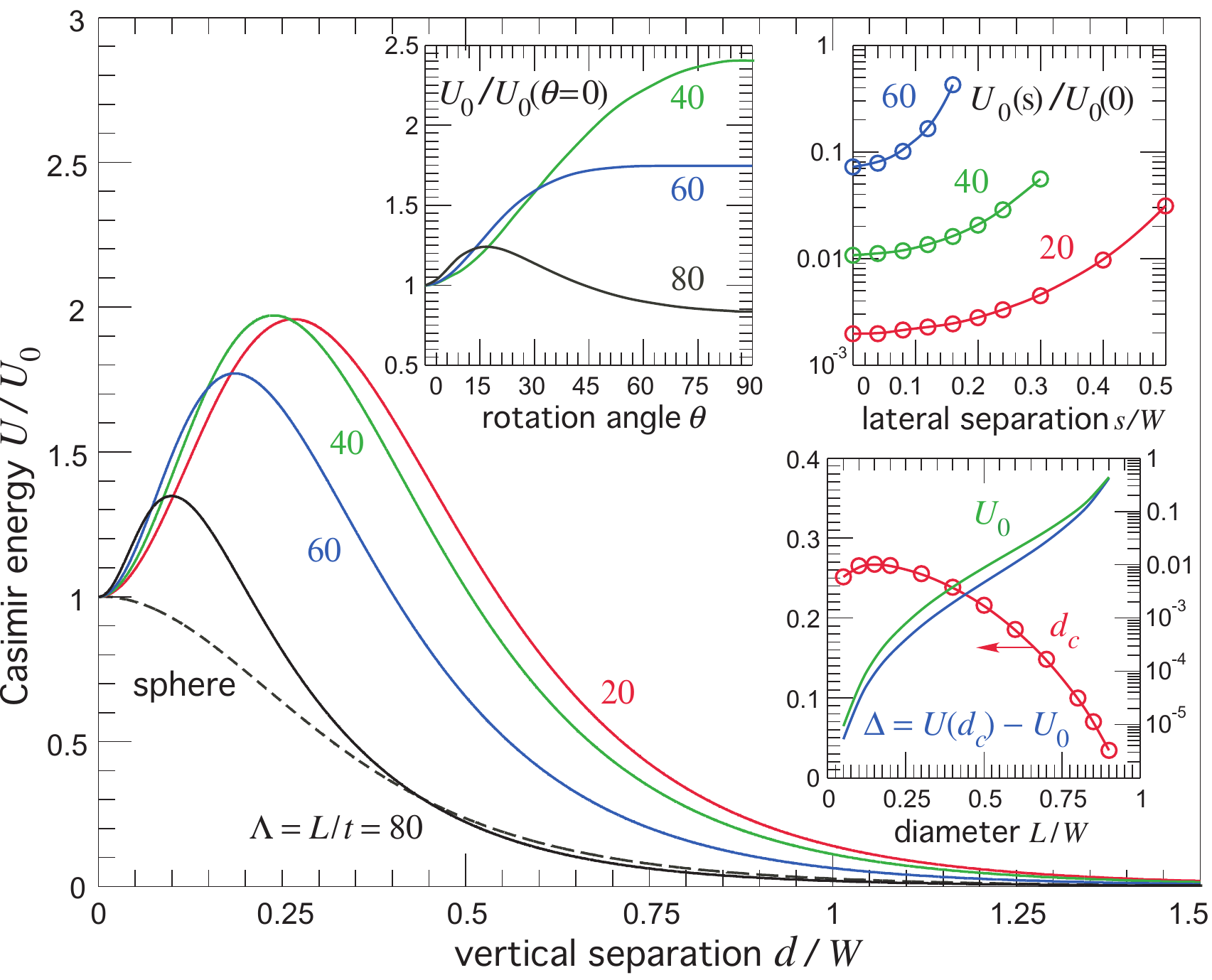}
\caption{Room-temperature Casimir energy $U$ as a function of $d$ for
  the disk--plate geometry of \figref{fig1}, but with the Au and PTFE
  materials interchanged. $U$ is normalized by $U_0$ and plotted for
  multiple values of $\Lambda$. (Top insets:) $U$ as a function of
  rotation angle $\theta$ (left) and lateral translations $s$ (right)
  for multiple $\Lambda$. (Bottom inset:) unstable equilibrium
  separation $d_c$ (red circles), along with $U_0$ (green line) and
  $\Delta = U(d_c)-U_0$ (blue line), normalized by $k_B T\approx
  25$m$eV$, as a function of $L$. Both $d_c$ and $L$ are normalized by
  $W$.}
\label{fig:fig3}
\end{figure}

\Figref{fig3} shows the ratio $U/U_0$ for the same geometry of
\figref{fig1} but for the ``reciprocal'' situation where the Au and
PTFE materials are interchanged (corresponding to a Au disk above a
PTFE plate). As before, the insets explore the stability of the system
with respect to rotations and lateral translations of the disk. In
this case, the polarizability of the disk is largest along the lateral
($x$--$y$) directions, and hence the relevant equation describing the
resulting CP interaction is the top equation of
\eqref{Udisk}. However, unlike the previous case, here it is the
$G_{xx}$ and $G_{yy}$ components of the DGF (and not $G_{zz}$) that
exhibit non-monotonicity, leading again to a meta-stable equilibrium
at $d=0$, albeit with slightly smaller $U$ and significantly larger
non-monotonicity for the same $\Lambda$. Essentially, the index
contrast between the Au disk and the fluid is orders of magnitude
larger than for a PTFE disk, leading to larger polarization
anisotropies. Unfortunately, the enhanced anisotropy comes at a price:
First, the small index contrast between the plate and the fluid
results in a smaller $U$, a consequence of the larger contribution of
the plate area. Second, the transition in the preferred orientation of
the disk from $\theta=0\to 90^\circ$ occurs at smaller $\Lambda$. The
large polarization anisotropy of the Au disk also means that the
potential trap is not very sensitive to the disk thickness. Fixing
$h=10$nm and $L=0.6W$, we find that $\Delta \to 2 \Delta$ and $U_0 \to
5 U_0$ as $t$ is increased from $t=10\mathrm{nm}\to 100$nm
(corresponding to a decrease in $\Lambda$ from $\Lambda=60\to 6$). On
the other hand, we find that $\Delta$ is very sensitive to changes in
the PTFE plate thickness. Fixing $t=10$nm and $L=0.6W$, we find that
$\Delta \to 0$ rapidly as $h$ is increased from $h=10\mathrm{nm}\to
100$nm. The situation is reversed in the reciprocal configuration of
\figref{fig1}, in which case the trap is sensitive to the disk
thickness and not the plate thickness.

The system described in this work constitutes a promising platform to
investigate two unusual geometry-induced Casimir phenomena: a
violation of the DLP condition of fluid repulsion between planar
bodies, and the stable suspension of two bodies. At room temperature,
the resulting ``Casimir trap'' has a depth on the order of $k_B T$,
which, unlike the case of a needle in vacuum~\cite{LevinMc10}, allows
for simpler (and more varied) experimental verification of this
phenomenon. It could also open new horizons for technological
applications where passive suspension is relevant. We believe that
even more pronounced effects should arise in other geometries and
material configurations. For instance, stronger potential traps might
be obtained by designing the shapes of the suspended bodies to exhibit
larger polarization anisotropy, a subject of future work.

This work was supported by DARPA Contract No. N66001-09-1-2070-DOD, by
the AFOSR Multidisciplinary Research Program of the University
Research Initiative (MURI) for Complex and Robust On-chip
Nanophotonics, Grant No. FA9550-09-1-0704, and by the U.S. Army
Research Office under contracts W911NF-07-D-0004 and W911NF-13-D-0001.


\end{document}